\documentclass[10pt,letterpaper]{article}
\usepackage[top=0.85in,left=1in,footskip=0.75in,marginparwidth=2in]{geometry}

\usepackage[utf8]{inputenc}

\usepackage{cite}
\usepackage{amsmath}
\usepackage{gensymb}
\usepackage{xfrac}
\usepackage{soul}
\usepackage{xcolor}

\usepackage{nameref}
\usepackage[hidelinks]{hyperref}

\usepackage{microtype}
\DisableLigatures[f]{encoding = *, family = * }

\usepackage{changepage}

\usepackage[aboveskip=1pt,labelfont=bf,labelsep=period,singlelinecheck=off]{caption}

\makeatletter
\renewcommand{\@biblabel}[1]{\quad#1.}
\makeatother

\usepackage{lastpage,fancyhdr,graphicx}
\usepackage{epstopdf}
\fancyheadoffset[L]{2.25in}
\fancyfootoffset[L]{2.25in}

\usepackage{color}

\definecolor{Gray}{gray}{.25}

\usepackage{graphicx}

\usepackage{sidecap}

\usepackage{wrapfig}
\usepackage[pscoord]{eso-pic}
\usepackage[fulladjust]{marginnote}
\reversemarginpar

\begin{document}
\vspace*{0.35in}

\begin{center}
{\Large
\textbf\newline{Optoelectronic properties of silver doped copper oxide thin films}
}
\newline
\\
Vishal Mohade\textsuperscript{1},
Krishna Kumar\textsuperscript{1}
Parasuraman Swaminathan\textsuperscript{1,2*}

\bigskip
\textsuperscript{1}Electronic Materials and Thin Films Lab,
Dept. of Metallurgical and Materials Engineering, \\
Indian Institute of Technology, Madras, Chennai, India \\
\textsuperscript{2}Centre of Excellence in Ceramics Technologies for Futuristic Mobility,\\
Indian Institute of Technology Madras, Chennai, India
\\

\bigskip
*Email: swamnthn@iitm.ac.in

\end{center}

\section*{Abstract}
Thin films have found a wide variety of applications because of the substantial improvement in their properties as compared to bulk metals. Metal oxide thin films are increasingly being used in various fields and are especially important in functional applications. They can be either $p$- or $n$-type in nature depending on the materials, dopants, and preparation route. Copper oxide is an example of a $p$-type metal oxide, which finds application in solar cells, photo-electrochemical cells, gas sensors, supercapacitors, and thermoelectric touch detectors. Both copper (I) and copper (II) oxides can be grown with the lower valence state oxide stable at low temperature and the higher valence state obtained by annealing at higher temperatures. In this work, we modify the optical and electrical properties of copper oxide thin films, by doping of silver through a thermal evaporation process route. Copper is thermally evaporated onto the substrate and silver is co-evaporated during this process. The films are then annealed in ambient under various conditions to obtain copper oxide. Structural and functional comparison is made between undoped and silver doped copper oxide thin films, prepared under the same conditions. Thermal evaporation is a simple route for obtaining doped metal oxides and the process can be extended to a variety of other systems as well. 

\bigskip

\noindent \textbf{Keywords:} Copper oxide; Thermal evaporation; Optoelectronic properties; Electron microscopy; Silver doping

\bigskip
\newpage
\section {Introduction}
Copper oxide is a well-studied material because of the abundance of copper in nature, its $p$-type conductivity, easy synthesis by oxidation of copper, good optical and electrical properties, and non-toxic nature. There are various methods to produce copper oxide thin films viz. thermal evaporation \cite{Choudhary2016growth, Figueira2017optimization, Figueiredo2013p, Papadimitropoulos2005deposition}, magnetron sputtering\cite{Murali2015synthesis}, electrodeposition\cite{Dhanasekaran2012electrochemical, Mathew2001temperature, Messaoudi2016structural, Paracchino2011highly}, chemical vapor deposition \cite{Barreca2007cvd}, chemical bath deposition \cite{Dubal2010fabrication, Lin2015copper, Sultana2017chemical}, plasma sputtering \cite{Li2017reactive}, molecular beam epitaxy, DC reactive
sputtering\cite{Alajlani2017characterisation}, RF reactive sputtering\cite{Ghosh2000deposition, Hartung2015assessing}, ion beam sputtering \cite{Horak2016thin}, sol-gel \cite{Raship2017effect, Ray2001preparation}, and pulsed laser deposition\cite{Chand2014structural}.

There have been several studies to determine the properties of copper oxide thin films. Pure copper film on annealing is found to form different oxides at different annealing temperatures. The first oxide is cuprous oxide (Cu$_2$O), which starts forming in range of 200-250 \degree C and exhibits and optical bandgap of 2.0-3.0 eV, with cubic crystal structure of lattice parameter 0.427 nm. The higher valence, cupric oxide (CuO) starts forming above 300 \degree C and has a optical bandgap of 1.2-1.7 eV. CuO shows monoclinic crystal structure and both oxides show $p$-type
conductivity \cite{Choudhary2016growth, Murali2015synthesis, Papadimitropoulos2005deposition}. These oxides shows promising applications in gas sensors\cite{limkrailassiri2013copper, nayan2016correlation, navale2017thermally}, solar cells\cite{Alajlani2017characterisation, roos1983properties, daoudi2019study}, thermoelectric touch sensors\cite{Figueira2017optimization}, thin film transistors\cite{Figueiredo2013p}, and supercapacitors\cite{Dubal2010fabrication}, to name a few. Thermal evaporation produces very uniform films with no porosity, and high adhesion. It is a non-toxic and line of sight process.

The Cu-Ag phase diagram \cite{kawecki2012fabrication} shows a eutectic obtained at 71.9 wt. \% Ag. For temperatures below 400 \degree C and very low concentration of Ag there is negligible mixing or solubility of Ag in Cu. The pressure vs. temperature
diagram for the Cu-O system \cite{schramm2005thermodynamic} shows that at a pressure greater than 100 Torr and temperatures higher than 600 \degree C, copper oxide exists as CuO. Papadimitropolous \textit{et al.} deposited copper oxide thin films using thermal vacuum evaporator on a silicon substrate \cite{Papadimitropoulos2005deposition}. On oxidation, the film showed cuprous oxide formation at 225 \degree C, copper
silicides were also seen at this temperature. At 280 \degree C, Cu$_2$O amount starts to decrease and CuO peaks appear, which then converts to only CuO peaks at 350 \degree C. CuO forms because the Gibbs free energy for the oxidation of Cu$_2$O to CuO at a
temperature of 200 \degree C is -3.73 kcal$/$mol \cite{musa1998production}. Chaudhary
\textit{et al.} justified it as the supply of thermal energy to the cubic Cu$_2$O should cause higher ionicity and smaller grain size to transform into lower symmetry and larger grain size \cite{choudhary2018oxidation}. Figuieredo
\textit{et al.} observed similar peaks of Cu$_2$O (111) from 250 to 300 \degree C and CuO (11-1) above 300 \degree C by using e-beam evaporation of pure copper sample on glass slide followed by annealing \cite{Figueira2017optimization}. Thus, the overall conversion can be written as 
\begin{equation}
    Cu \rightarrow (Cu\;+\;Cu_2O) \rightarrow Cu_2O \rightarrow (Cu_2O + CuO) \rightarrow CuO
\end{equation}

In this work we attempted addition of silver nanoparticles on the surface of a copper thin film without breaking vacuum. We used a thermal evaporation process route to ascertain efficient film morphology for study of the functional properties. The as deposited samples were then annealed to obtain oxides. The study of morphology helped us to understand grain development with temperature. A variety of characterization tools were used to study the effects of silver addition. The optical properties showed the bandgap values of copper oxide similar to others. Similar work has been attempted by only one other researcher, who used microwave annealing process to study optical properties of silver doped copper oxide thin film \cite{das2013structural}. Li-doped copper oxide thin films also shows a decrease in band gap with Li
concentration \cite{chand2014structural}. The doping process for copper oxide thin film has not been adapted elsewhere.
The effect on electrical properties because of doping copper oxide have not yet been reported. This work will give a brief information about changes in resistance and carrier concentration by silver doping. The effect of temperature on resistance of copper oxide thin film can also help to develop a temperature sensor. These properties were studied elsewhere with computational modelling using density functional theory \cite{absike2019electronic}.

\section {Materials and methods}

Copper thin films were deposited on microscopic glass slides using high vacuum coating unit (Model HPVT 303) by Hydro Pneo Vac Technology. Prior to deposition, the glass slides were cut to dimension of 2.5 $\times$ 2.5 cm$^2$. Then they were cleaned in steps of 1 \% soap solution, deionized water, ethanol and deionized water again by ultra-sonication for 15 min in each solution. These were dried by wiping with lint free cloth. The deposition was performed at high vacuum level ($7 \times 10^{-6}$ mbar). The rate of deposition was maintained for all depositions to be ~ 0.14  nm s$^{-1}$. Thin film thickness was monitored with a quartz crystal microbalance. The substrate was kept at room temperature and no external heating was provided during deposition, though there was a rise in temperature of the substrate during thermal evaporation. Silver thin film of thickness 1 nm were deposited on top of the pure copper film, without breaking vacuum. The deposited thin film was then annealed at temperatures from 150 to 450 \degree C at an interval of 100 \degree C. The annealing was performed in a muffle furnace in ambient atmosphere. The heating rate was maintained at 5 \degree C/min and the holding time was 2 h. Both pure copper and silver doped copper films were annealed under the same conditions.

Grazing incidence x-ray diffraction (GIXRD) measurements were performed on a Rigaku Smartlab XRD machine with 9 kW
rotating anode X-ray source. The configuration includes Cu as the source material and Ge monochromator. The incidence angle was maintained at 1 \degree and $2\theta$ was varied at the
scanning rate of 0.01 \degree per s. The analysis was done using XPert Highscore pro.cUV-Visible spectroscopy was performed on a Jasco V-730 spectrophotometer, with halogen and
deuterium lamps as sources. The wavelength range for measurement was 200 – 1100 nm. Raman spectroscopy was performed using NdYAG laser of wavelength 532 nm on WiTech alpha 300 confocal Raman microscope. The accumulation time was kept at 5 s. The
objective used was 20 $\times$. For photoluminescence spectroscopy (PL) Jasco FP 6300 was used. The excitation wavelengths used were 320 nm and 450 nm. A DC powered 150 W Xe lamp was used as a
source. The accumulation and integration time were set at 10 s respectively. Electrical transport measurement was carried out in Van der Pauw setup on DynaCool PPMS by Quantum design at room temperature. All measurements were done at 300 K. The contacts
were made using silver paint adhesive and copper wire. The sample dimensions were maintained at 4 mm $\times$ 4mm ($L \times W$).
$I-V$ measurements were performed in a four-probe setup on a source meter, Keysight precision measurement system. The contacts were made using silver epoxy adhesive and copper wires. Voltage range was kept as -10 V to 10 V. $I-V$ measurement with temperature (Seebeck coefficient) was done on Cascade Microtech summit 12000 AP. The temperature range was from room temperature to 120 \degree C in steps of 10 \degree C and ETC 200L (Espec corp.) thermos chuck was used to monitor temperatures. The voltage range was -10 to 10 V.
Sheet resistance was measured on Jandel mode RM 3000 using a four probe configuration. The equipment was calibrated and zeroed before taking readings. The current was automatically set by equipment within range of 10 nA to 99.99 mA. Scanning electron microscopy was done on FEI Quanta 400 high resolution scanning electron
microscope. The detector used was Everhart Thornley detector (ETD).
Transmission electron microscopy was performed on FEI Tecnai 12 electron microscope.

\section {Results and Discussion}
\subsection{Structural characterization}
\begin{figure}
    \centering
    \includegraphics[width=16cm]{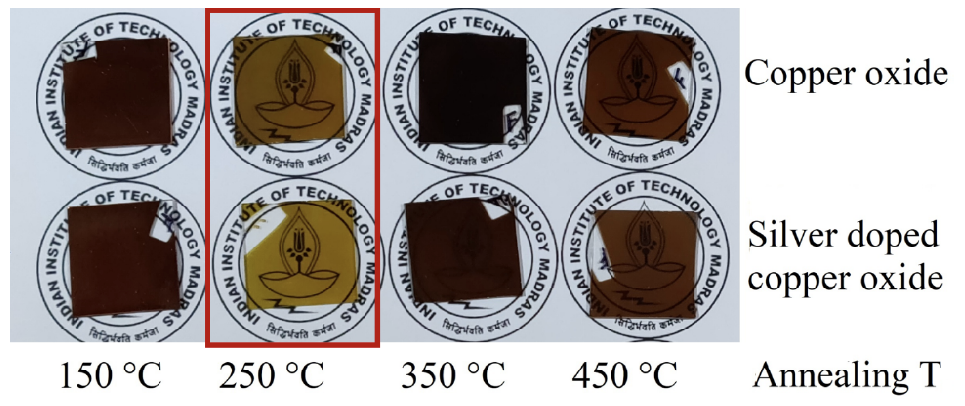}
    \caption{Comparison of optical images of annealed pure copper oxide and silver doped copper oxide thin films.}
    \label{fig:Fig1}
\end{figure}

On annealing of samples, visually, we can clearly deduce that the sample with 250 \degree C annealing temperature was the most transparent. Annealing at lower temperatures, 150 \degree C, produced samples which were opaque, while at the highest temperature of 450 \degree C, the annealed sample was more transparent than 350 \degree C annealed sample. The optical images are summarized in figure \ref{fig:Fig1} and the samples are kept above IITM logo for showcasing their transparency. 

\begin{figure}
    \centering
    \includegraphics[width=14cm]{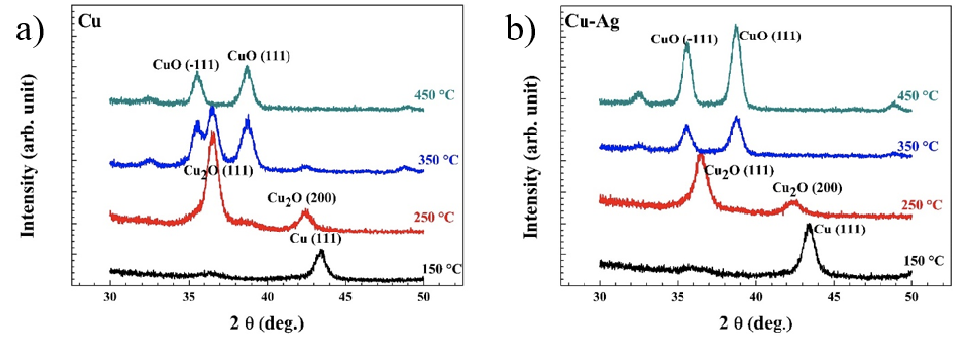}
    \caption{GIXRD of a) undoped copper oxide b) silver doped copper oxide thin film at different annealing temperatures.}
    \label{fig:Fig2}
\end{figure}

To identify the phases, XRD studies were carried out on copper oxide and silver doped copper oxide thin films. The XRD patterns were measured for thin films annealed at temperatures from 150 to 450 \degree C. GIXRD results, presented in figure \ref{fig:Fig2} (a) and (b), showed the presence of copper peaks in 150 \degree C annealed sample (ICDD number: 01-085-1326), indicating incomplete oxidation at this temperature. Samples annealed at 250 \degree C showed peaks at 36.5 \degree corresponding to Cu$_2$O (111) and at 42.2 \degree corresponding to Cu$_2$O (200) (ICDD number: 05-667). Samples annealed at 350 \degree C showed peaks at 35.5 \degree and 38.7 \degree corresponding to CuO [ICDD number: 45-0937] and samples annealed at 450 \degree C showed 35.5 \degree and 38.7 \degree peaks. At 150 \degree C, a copper peak appears at 42.4 \degree, from Cu (111) plane. The data obtained matches results are obtained by Choudhary \textit{et al.}\cite{Choudhary2016growth}.

The crystallite size variation with temperature and doping can be seen in table \ref{tab:Table1}. The crystallite size was obtained using Hall–Petch equation. For undoped copper oxide thin films the crystallite size decreases with annealing temperature, whereas for Ag doped copper oxide it increases. Ag doping has decreased the crystallite size in films annealed at 150 and 250 \degree C, when compared to undoped Cu at these temperatures. This is due to higher affinity of copper towards oxygen and the preferential oxidation of copper that occurs in presence of silver \cite{das2013structural}. 
\begin{equation}
    Ag_2O\,(s) \; + \; Cu\,(s) \rightarrow CuO\,(s) \; + \; 2Ag \,(s)
\end{equation}

\begin{table}[h]
    \centering
    \caption{Variation of crystallite size with temperature.  \\}
    \begin{tabular}{|c|c|c|}
    \hline
    \textbf{Annealing}&\textbf{Crystallite size of} & \textbf{Crystallite size of}  \\
    \textbf{temperature (\degree C)}&\textbf{undoped copper oxide (nm)} & \textbf{silver doped copper oxide (nm)} \\
    \hline
    150 & 33.7 & 5.2 \\ \hline
    250 & 13 & 6.3 \\ \hline
    350 & 8.3 & 18.8 \\ \hline
    450 & 10.6 & 15.2 \\    
         \hline
    \end{tabular}
    \label{tab:Table1}
\end{table}

\begin{figure}
    \centering
    \includegraphics[width=12cm]{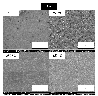}
    \caption{SEM images of pure copper oxide thin film annealed at 150, 250, 350, and 450 \degree C.}
    \label{fig:Fig3}
\end{figure}

\begin{figure}
    \centering
    \includegraphics[width=12cm]{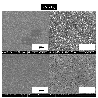}
    \caption{SEM images of Ag doped copper oxide thin film annealed at 150, 250, 350, and 450 \degree C.}
    \label{fig:Fig4}
\end{figure}

SEM was performed at magnification of 40000 $\times$ in secondary electron mode using a Everhart–Thornley detector. Thin films were sputtered with gold before measurement. Typical microstructure of the thin films, without and with silver, analyzed by SEM are shown figures \ref{fig:Fig3} and \ref{fig:Fig4} respectively. The inset at the bottom left shows the calibration scale to observe the grain size and top right shows the temperature of annealing. Both thin films annealed at 150 \degree C shows smoother film and very fine grain size and closed packed structure. The grain size increases with annealing temperature. Figure \ref{fig:Fig3} shows undoped copper oxide thin films annealed at different temperatures. The film morphology appears smooth and grain size distribution is uniform. Thin films annealed at 350 \degree C shows very high increase in grain size as compared to other films. The film
annealed at 450 \degree C shows presence of pores smaller than the grains. Ag doped SEM images, on the other hand, shows smooth and uncracked films. The grains appear well-defined in Ag doped copper oxide thin film (shown in figure \ref{fig:Fig4}) and the grain size increases with temperature, which is similar to the undoped copper oxide thin film. The grains agglomerated to form bigger grains, therefore grain size variation in the 450 \degree C annealed film looks non uniform. A crack is also observed around bigger grains. This crack is due to connecting of adjoining pores. The pore
size observed is comparable to the grain size. The 350 \degree C film shows more layered structure and decreased pore size. On comparing to undoped copper oxide thin film annealed at 450 \degree C in figure \ref{fig:Fig3}, the Ag doped thin film shows aggregated grains with more defined shape of top layer.

\begin{figure}
    \centering
    \includegraphics[width=16cm]{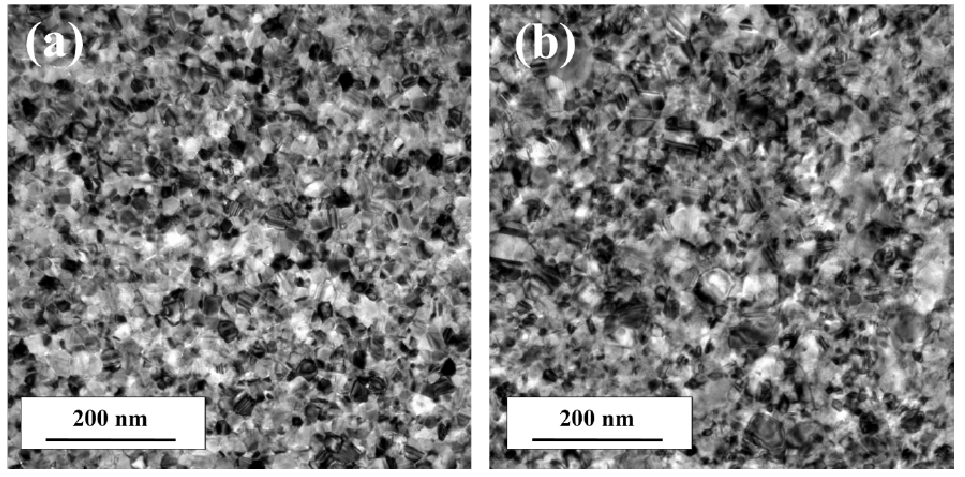}
    \caption{TEM image of a representative sample of a) pure copper and b) silver doped copper}
    \label{fig:Fig5}
\end{figure}

To get closer look at the grain structure on the addition of silver to copper thin films, representative TEM images of the as deposited copper and silver doped copper thin films are shown in figure \ref{fig:Fig5}. Figure \ref{fig:Fig5} (a) shows a uniform
grain size for pure copper, whereas in figure \ref{fig:Fig5} (b) the grain size is non uniform and there are some grains which are distributed randomly with larger size as compared to rest of the grains and the smaller grains shows similar size as in fig \ref{fig:Fig5} (a). By comparison between (a) and (b) we
can conclude that the bigger grains are of Ag doped copper. 

\subsection{Optical properties}

\begin{figure}
    \centering
    \includegraphics[width=16cm]{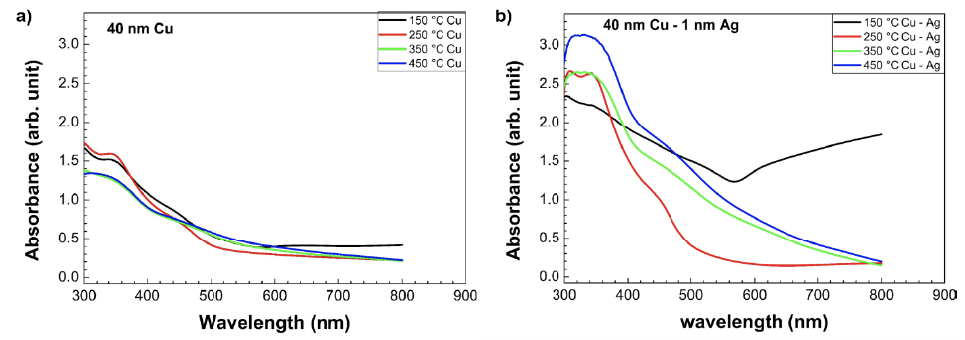}
    \caption{Variation of absorbance with wavelength for a) pure copper oxide thin film and b) silver doped copper oxide thin film.}
    \label{fig:Fig6}
\end{figure}

\begin{figure}
    \centering
    \includegraphics[width=16cm]{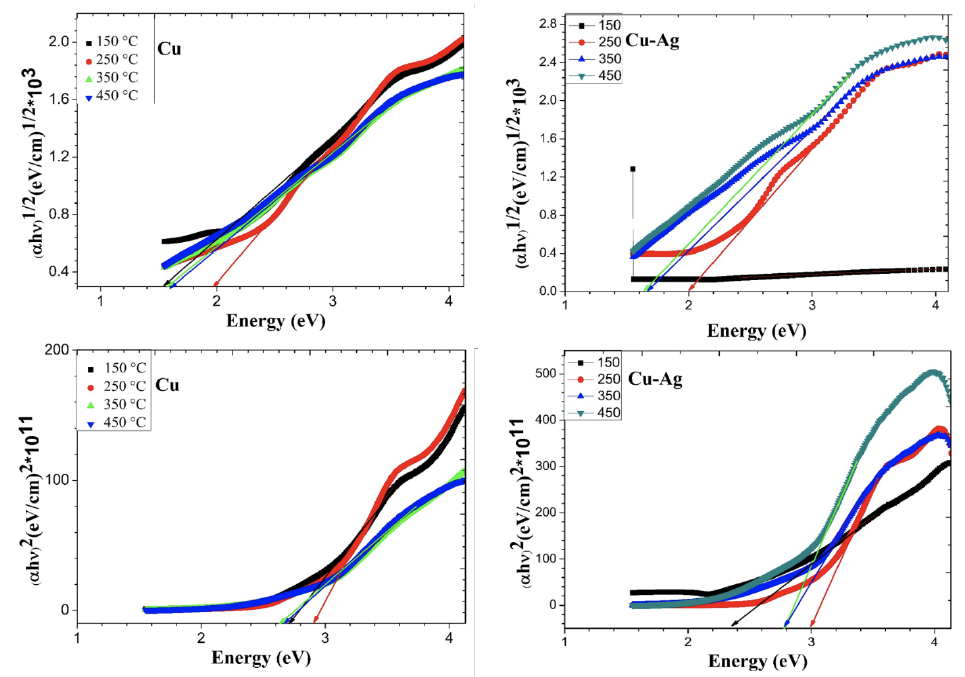}
    \caption{Tauc plot for undoped and Ag doped Copper oxide thin films.}
    \label{fig:Fig7}
\end{figure}

Figure \ref{fig:Fig6} shows the absorbance plot for both doped and undoped copper oxide thin films. The absorbance is high in the UV region due to the opaque nature of the thin film. Silver doped thin
film shows higher absorbance in blue region as compared to undoped copper oxide. This may be due to the nature of noble metals such as silver. They absorb light due to the transition of electrons between unoccupied hybridized $sp$ states and occupied $d$ states. Absorbance decreases with increase in wavelength till 500 nm (blue green region) for undoped thin film. A similar behavior is shown by silver doped copper oxide annealed at 250 \degree C. The optical band gap of copper oxide thin films can be calculated using Tauc equation and figure \ref{fig:Fig7} shows the Tauc plot for determining the band gap of CuO and Ag doped CuO thin
films. Band gap values can be seen from the extrapolation of the linear portion of to the energy axis of the Tauc plot (arrow pointing on $X$-axis). Cu$_2$O has a direct band gap of 2.91 $\pm$ 0.3 eV and CuO has indirect band gap of 1.5 $\pm$ 0.3 eV. The data obtained is similar to those obtained in literature \cite{Choudhary2016growth, Figueiredo2013p, Figueira2017optimization}. The band gap of silver doped Cu$_2$O is 3.04 $\pm$ 0.3 eV and CuO is 1.70 $\pm$ 0.3 eV.

\begin{figure}
    \centering
    \includegraphics[width=16cm]{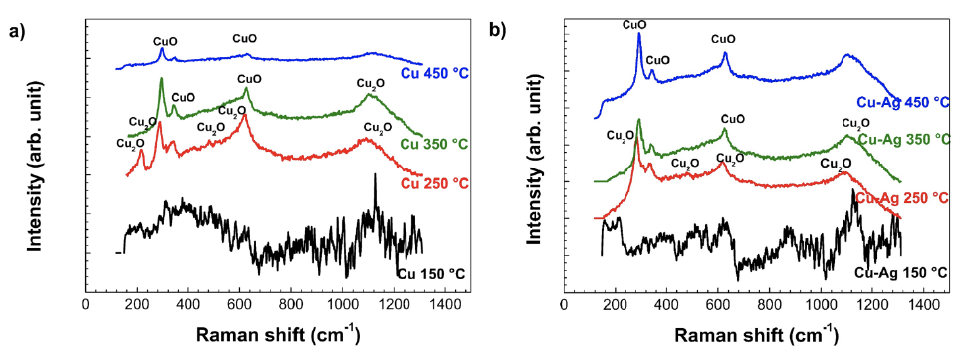}
    \caption{Raman spectra of annealed a) undoped copper oxide and b) silver doped copper oxide thin films.}
    \label{fig:Fig8}
\end{figure}

Figure \ref{fig:Fig8} shows Raman spectra of annealed copper oxide thin films. Among nine optical modes of CuO, three are Raman active ($A_g$+ 2 $B_g$). The peaks 298, 342 and 627 cm$^{-1}$ corresponds to CuO and the peaks 143, 210, and 617 cm$^{-1}$ correspond to Cu$_2$O. 450 \degree C Ag doped copper oxide thin film shows all three CuO peaks only. On the other hand, other temperatures
show the presence of Cu$_2$O peaks more dominantly. The silver doped copper oxide thin film annealed at 250 \degree C shows the greatest intensity of peaks at 143 and 210 cm$^{-1}$. This can be attributed to the presence of Cu$_2$O. Raman data confirms the results obtained by XRD. Undoped and doped copper oxide films annealed at 150 \degree C shows only one peak corresponding to Cu$_2$O at 1100 cm$^{-1}$. This signifies the presence of very thin oxide layer, although no other peaks are available. This can suggest that the oxide formation has not started properly at 150 \degree C.

\begin{figure}
    \centering
    \includegraphics[width=16cm]{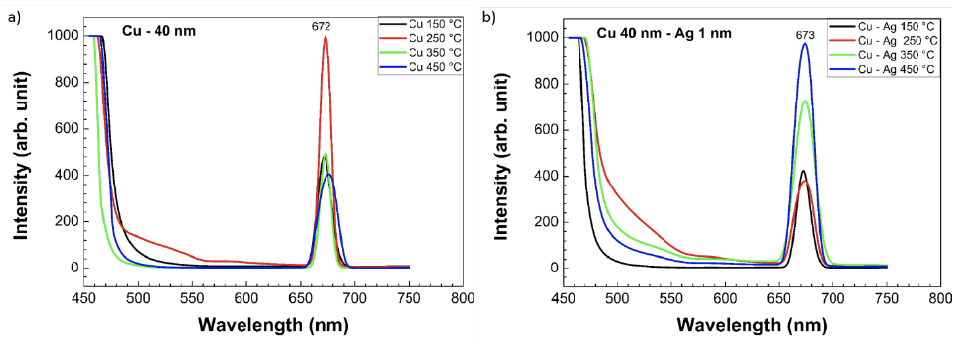}
    \caption{Photoluminescence (PL) spectra of a) undoped copper oxide thin film and b) silver doped copper oxide thin film with an excitation wavelength 450 nm.}
    \label{fig:Fig9}
\end{figure}

Photoluminescence spectra were also recorded for the undoped and doped copper oxide thin films. The film annealed at 250 \degree C
shows peaks at 398 nm (near UV emission) and 496 nm (corresponding to cyan emission). The peaks at 540 and 578 nm correspond to green and yellow emissions respectively. The presence of high yellow peaks indicates excess oxygen whereas green emissions indicates oxygen deficiency \cite{mitra2001synthesis, djurivsic2006green}. Both doped and undoped copper oxide thin films annealed at 450 \degree C exhibit emission in UV range i.e. 364 nm, which can be correlated to band gap transition in CuO \cite{djurivsic2006green}. The green emissions are also associated with the presence of surface defects and transition of carriers from near conduction band (oxygen vacancies) to deep valence band (Cu vacancies). Figure \ref{fig:Fig9} shows PL spectra for undoped and doped copper oxide thin films at an excitation wavelength 450 nm. The graph shows a single red emission, this can be due to neutral and single ionized oxygen transitions or deep level emission related to presence of defect \cite{el2005influence, vanheusden1996correlation}. 

\subsection{Electrical properties}

Electrical transport measurements were performed with Van der Pauw method to obtain carrier concentrations. It shows the resistance value obtained for undoped copper oxide thin films annealed at 250 and 450 \degree C are 12-20 M$\Omega$ and 2.8 M$\Omega$ respectively. For silver doped thin film, annealed at 450 \degree C, the corresponding value was 22-30 k$\Omega$. The carrier concentration values for undoped Cu$_2$O thin film were in range of $10^{15}$ cm$^{-3}$ and for doped CuO in range of $10^{16}$ cm$^{-3}$. The nature of the Hall coefficient shows $p$-type conductivity in copper oxide thin films \cite{Figueiredo2013p, Figueira2017optimization, Murali2015synthesis}.

Table \ref{tab:Table2} shows the sheet resistance values obtained by four probe measurement. The films annealed at 150 \degree C showed same resistance as copper because of negligible surface oxidation. The resistance is very low for undoped copper oxide thin films as compared to silver doped copper oxide thin films. The increase in resistance is due to increased scattering centers by
the addition of silver.

\begin{table}[h]
    \centering
    \caption{Sheet resistance values obtained by four probe measurement.  \\}
    \begin{tabular}{|c|c|c|}
    \hline
    \textbf{Annealing}&\textbf{Undoped copper oxide} & \textbf{Silver doped copper oxide}  \\
    \textbf{temperature (\degree C)}&($\Omega/sq.$) & ($\Omega/sq.$)\\
    \hline
    150 & 1.2 & 1.1 \\ \hline
    250 & $4 \times 10^6$ & $66 \times 10^6$ \\ \hline
    350 & $10 \times 10^6$ & $24.7 \times 10^6$ \\ \hline
    450 & $1.3 \times 10^6$ & $19 \times 10^6$ \\    
         \hline
    \end{tabular}
    \label{tab:Table2}
\end{table}

The $I-V$ curves were also obtained for undoped and silver doped copper oxide thin films. The undoped copper oxide thin film annealed at 450 \degree C shows non-linear behavior, while the other samples showed Ohmic behavior. The resistance values obtained were similar to those obtained by four-probe data. The temperature dependence of resistance was also measured and the values are listed in table \ref{tab:Table3}. The table shows that the resistance decreases with increase in temperature for all the
annealed thin films. The drop in the resistance for silver doped thin film annealed at 250 \degree C is very drastic, but the initial resistance value is very high which also can be seen the from table. This decrease is not regular for undoped copper oxide thin film annealed at 250 \degree C. The decrease for silver doped thin film annealed at 450 \degree C is gradual with temperature, which makes it a great choice for temperature sensor application.

\begin{table}[h]
    \centering
    \caption{Resistance variation with temperature for select annealed copper oxide thin films.  \\}
    \begin{tabular}{|c|c|c|c|}
    \hline
    \textbf{Temperature}&\textbf{Resistance of} & \textbf{Resistance of} & \textbf{Resistance of}  \\
    \textbf{(\degree C)}&\textbf{Cu 250 \degree C} (M$\Omega$) & \textbf{Cu-Ag 250 \degree C} (M$\Omega$) & \textbf{Cu-Ag 450 \degree C} (M$\Omega$)\\
    \hline  
    22 & 33.3 & 203 & 14. 7 \\ \hline
    50 & 9.05 & 92.7 & 9.37 \\ \hline 
    60 & 17.9 & 69 & 8.02 \\ \hline
    70 & 6.66 & 52.3 & 6.71 \\ \hline
    80 & 8.87 & 42.1 & 6.1 \\ \hline
    90 & 5.95 & 30.5 & 5.49 \\ \hline
    100 & 6.34 & 23.1 & 4.61 \\
         \hline
    \end{tabular}
    \label{tab:Table3}
\end{table}

\section{Conclusion}

Copper oxide thin film annealed at 250 \degree C shows the highest optical transparency. XRD studies shows the presence of Cu$_2$O phase in both undoped and silver doped copper oxide thin films
annealed at 250 \degree C. CuO phase starts forming above 330 \degree C and hence it is visible in XRD pattern of thin films annealed at 350 and 450 \degree C. Raman spectroscopy also confirm the results obtained by XRD. On increasing the annealing temperature, the grain size increased. Silver doped copper oxide thin films showed larger band gap than undoped copper oxide thin
films. The blue and green emission showed deficiency or excess of oxygen in thin films, and red emission showed transition of single and ionized oxygen. Copper oxide thin films showed $p$-type conductivity. Silver doped copper oxide thin films showed higher resistance than undoped copper oxide. The resistance value decreases with increase in temperature for all copper oxide thin films. The resistance variation of silver doped copper oxide thin film annealed at 450 \degree C with temperature showed a possible application as temperature sensor.

\section*{Acknowledgments}
Support from the Centre of Excellence in Ceramics Technologies for Futuristic Mobility (project number SB22231272MMETWO008702) is acknowledged. Electron microscopy (SEM and TEM) was carried out at the facilities available in the Dept. of Metallurgical and Materials Engineering, IIT Madras. Optical characterisation was performed at the facilities available at the Dept. of Physics, IIT Madras, while electrical characterisation was carried out at the Centre for NEMS and Nanophotonics, IIT Madras.


\bibliography{library}

\bibliographystyle{hunsrt}
\end{document}